\documentclass[11pt]{article}
\def\be{\begin{equation}}
\def\ee{\end{equation}}
\def\ba{\begin{eqnarray}}
\def\ea{\end{eqnarray}}

\def\ra{\rangle}

\def\lo{\longrightarrow}

\usepackage{epsfig}
\begin{document}
\begin{titlepage}
\vspace{4cm}
\begin{center}{\Large \bf Reply to \\ Comment on\\
''Quantum key distribution for d-level systems with generalized Bell
states''
}\\
\vspace{1cm} V. Karimipour\footnote{email: vahid@sharif.edu},\\
\vspace{1cm} Department of Physics, Sharif University of Technology,\\
P.O. Box 11365-9161,\\ Tehran, Iran
\end{center}
\vskip 3cm

\begin{abstract}
In a recent comment \cite{ch1} it has been claimed that an
entangled-based quantum key distribution protocol proposed in
\cite{zhang} and its generalization to d-level systems in \cite{v1}
are insecure against an attack devised by the authors of the
comment. We invalidate the arguments of the comment and show that
the protocols are still secure.
\end{abstract}

\end{titlepage}
In a recent comment \cite{ch1}, it has been argued that the
protocols of quantum key distribution proposed in \cite{zhang} and
its generalization to d-level systems \cite{v1}, are insecure
against a special type of attack. Since the comment has been
addressed only to the d-level scheme proposed in \cite{v1}, we use
in this reply  the language of $d-$ level systems, although we think
that the authors could have primarily addressed their comment to the
original two-level scheme proposed in \cite{zhang} (and cited in
\cite{v1}), in which case the whole discussion
including their own, would have been much simpler. \\

The basic idea of \cite{v1} is that Alice(a) and Bob(b) can share
a maximally entangled state like a generalized Bell state

\begin{equation}\label{bell1}
    |\psi\ra=\frac{1}{\sqrt{d}}\sum_{j=0}^{d-1} |j,j\ra_{a,b}\ ,
\end{equation}

as a carrier between themselves. At the origin, Alice(a) can
entangle the qudit $q$ of the key (k) to this carrier by a local
operation ( a generalization of Controlled NOT to d-dimension,
namely a Controlled mod-d addition), $|j,q\ra_{a,k}\lo
|j,j+q\ra_{a,k} $, producing the state

\begin{equation}\label{bell2}
    |\psi\ra=\frac{1}{\sqrt{d}}\sum_{j=0}^{d-1} |j,j,j+q\ra_{a,b,k}\ .
\end{equation}

During the transmission the value of the $q$ will be protected
from Eve, (since the density matrix of this key qudit will be
maximally mixed) and only Bob, at the destination, can extract the
value of $q$ by disentangling it from the carrier by the reverse
local
operation $|j,j+q\ra_{b,k}\lo |j,q\ra_{b,k} $.\\

In the original articles \cite{zhang, v1}, the possibility of
entanglement of Eve with the states already possessed by Alice and
Bob has been taken into account and methods for preventing her
from acquiring useful information have been devised. However the
authors of the comment \cite{ch1} show that Eve can perform
suitable operations to intercept only the odd-numbered bits
without being recognized by
Alice and Bob. We note in passing that a similar comment has been made by these authors \cite{ch2}
on another protocol \cite{v2} which has been replied in \cite{v3}. \\

To the original objection of the author of this reply, pointed to
in \cite{v3}, that intercepting a fixed known subset does not
imply insecurity, the authors of the comment only add a vague and
inexact argument at the end of their comment and claim that Eve
can stop and re-start her attack at any time she wants and
therefore she can intercept any random subset of the key bits
unknown to the
legitimate parties. \\

We show in this reply that the above claim in un-justified. The
reason is quite simple. Stopping and re-starting the attack by Eve,
requires disentangling and re-entangling herself with the carrier
without destroying the carrier in possession of Alice and Bob. In
fact, as shown in the original papers and reproduced in equations
(3-7) of the comment, this entangling process is a fairly delicate
process, since Eve should entangle herself with the carrier of Alice
and Bob, only by performing local operations on her own qudit(e) and
the key qudit(k) which is in transmission. This important point has
been simply overlooked by the authors of the comment and as we will
show here explicitly when Eve entangles herself with the carrier in
the first round, she can disentangle and re-entangle herself with
the carrier only in the odd rounds of
transmission. \\

The authors of the comment do not present any explicit calculation
as to how Eve can disentangle herself from the carrier and suffice
to say that she can do this by " parity switching operations"
whose meaning is unclear. To help the authors we explicitly show
how she can do this only in the odd rounds and strongly insist
that the authors show the way for the even rounds.\\

Consider again equations (3-7) of  the comment \cite{ch1}, which
show the state of Alice(a), Bob(b), the key qudit in
transmission(k) and Eve(e) at various stages of the first round.
Equation (4) is the initial state when Eve wants to start her
attack, (i.e. entangle herself with the carrier). She does this by
a controlled operation on the qudits $k$ and $e$, in the form
$$|j+q_1,0\ra_{k,e}\lo |j+q_1,j+q_1\ra_{k,e} $$  (equation 4 and 5).
After Bob extracts the key, Alice and Bob are left  with the
carrier (7) which is now entangled with Eve. Eve can disentangle
herself at stage 5, by the reverse operation
$$|j+q_1,j+q_1\ra_{k,e}\lo |j+q_1,0\ra_{k,e}.$$ She can do this in
any other odd round. However the authors of the comment, for the
purpose of their attack, change this carrier by the operation
shown in equation (8) of \cite{ch1}. The carrier at the beginning
of round 2 is given in equation (9) of \cite{ch1}. For the second
round, the corresponding states are those of equations (9-14) of
\cite{ch1}. The authors of the comment do not show how at any
stage of this round or any other even round, Eve can disentangle
herself from the carrier. Unless they show this explicitly, their
argument about the possibility of stopping and restarting the
attack and hence intercepting arbitrary subsequences of the key
qudits is incomplete and their conclusion about the insecurity of
our protocol is
invalid. \\

In conclusion we have shown that Eve can stop and re-start her
attack only at odd rounds of transmission and in this way can only
intercept the odd-numbered key qudits, which by no means implies the
insecurity of such a protocol \cite{v3}.

{}
\end{document}